\documentclass{aa}
\usepackage{graphicx}
\begin{document}
\headnote{Research Note}
\title{On the usefulness of finding charts}

\subtitle{Or the runaway carbon stars of the Blanco \& McCarthy field 37}

   \author{C. Loup \inst{1}
           \and
           N. Delmotte \inst{2,3}
           \and
           D. Egret \inst{2}
           \and
           M.-R. Cioni \inst{3}
           \and
           F. Genova \inst{2}
          }

   \offprints{C. Loup}

   \institute{Institut d'Astrophysique de Paris, CNRS UPR 341,
              98bis Bd. Arago, F--75014 Paris, France\\
              \email{loup@iap.fr}
   \and
              CDS, Observatoire Astronomique de Strasbourg, UMR 7550, 
              Universit\'e Louis Pasteur, F--67000 Strasbourg, France
   \and
              European Southern Observatory, ESO, Karl Schwarzschild Str.2,
              D--85748 Garching bei M\"unchen, Germany
%
             }

   \date{Received 19-04-02; accepted 04-02-03}

\abstract{We have been recently faced with the problem of cross--identifying
stars recorded in historical catalogues with those extracted from recent
fully digitized surveys (such as DENIS and 2MASS). Positions 
mentioned in the old catalogues are frequently of poor precision,
but are generally accompanied by finding charts where the interesting objects
are flagged. Those finding charts are sometimes our only link with the
accumulated knowledge of past literature. While checking the identification
of some of these objects in several catalogues, we had the surprise
to discover a number of discrepancies in recent works.The main reason
for these discrepancies was generally the blind application of the
smallest difference in position as the criterion to identify sources 
from one historical catalogue to those in more recent surveys. In this paper
we give examples of such misidentifications, and show how we were able
to find and correct them.We present modern procedures to discover and
solve cross--identification problems, such as loading digitized images of the
sky through the Aladin service at CDS, and overlaying entries from
historical catalogues and modern surveys. We conclude that the use
of good finding charts still remains the ultimate (though time--consuming)
tool to ascertain cross--identifications in difficult cases.
   }

   \maketitle
%

\section{Introduction}

  The question adressed in this Research Note is the following : can one
cross--identify old catalogues listing inaccurate coordinates with modern
ones providing good coordinates using automatic blind matches ? Of course one
can, one just has to write or borrow rather simple routines and
to digitize the old catalogue. But is the error rate acceptable ? 

   Here we give a piece of
answer, based on the cross--identification of the carbon star catalogue
of Blanco \& McCarthy (1990, acronym LMC--BM) with the one of Kontizas et
al. (2001, acronym [KDM2001]), as well as with the 
DENIS Catalogue towards
the Magellanic Clouds (Cioni et al. 2000, acronym DCMC). The sensitivity
of both surveys is similar, but Blanco \& McCarthy were limited to 49 circular
regions of $\sim 0.12^{\circ^2}$, while Kontizas et al. have surveyed
the whole LMC. From the restricted point of
view of this article, the main difference between both surveys is the
astrometric accuracy, $\sim 1\arcsec$ for Kontizas et al., 
and better than $17\arcsec$ in Blanco \& McCarthy. Moreover,
the way the astrometry was performed in Blanco \& McCarthy involved much
more manual work than in Kontizas et al., leading to a much higher risk
of human errors, as will be seen in Sect.~2. 
Kontizas et al. have matched (automatically) 
both catalogues and present the results in their Table 3. During the course
of a more general work about cross--identifications in the LMC, 
we have checked some of their results by looking
at the finding charts. This led us to cross--identify the whole Blanco \&
McCarthy catalogue in the same ``old--fashion'' way, and to compare it
with automatic matches, as presented in Sec.~3. Sec.~4 lists a few
additional errata. Short conclusions are given in Sect.~5.

\section{The case of the Blanco \& McCarthy field 37}

   The case of the Blanco and McCarthy field number 37 illustrates very
well the risk of cross--identifications in a ``blind way'', that is, 
on the basis of poor coordinates only. We would like to suggest the reader
to visit the Centre de Donn\'ees Astronomiques de Strasbourg (CDS,
http://cdsweb.u--strasbg.fr), 
then the Aladin image facility, to type in the coordinates of,
for instance, LMC--BM 37--20 (05 43 42 --70 27.5, J2000), 
and to load the ESO MAMA or AAO DSS2 digitized
R image. At first view one does not recognize the field of
the finding chart. At second view neither. The experience can be made
with any star of the field 37, the field centered on the coordinates
never corresponds to the one of the finding chart. 
However, if one looks at, for instance, 
LMC--BM 37--24 (05 44 03 --70:26.1, J2000),
one can recognize the field thanks to two bright stars, both very far
away from the expected location, more precisely about 50s 
($4\arcmin$) towards the West and $2\arcmin$ towards the South. 
It turns out that all the positions
of the LMC--BM stars in the field 37 are erroneous by about $4.5\arcmin$,
as well as the position of the field center. 
(The reason for this error remains unknown).

   Kontizas et al. find 13 cross--identifications in this field, 29\% of
the total number of C stars found by Blanco \& McCarthy. 
As the shift of the position is about $4.5\arcmin$, while 
the search radius of Kontizas et al. was about $1\arcmin$,
it is clear that the 13 cross--identifications given by Kontizas et al. 
must all be erroneous. They are just random associations. This problem 
has some consequences on the cross--identifications with field 42 as well,
because it is located very close to field 37. 
Five cross--identifications
were missed because the corresponding [KDM2001] star was already associated
to an LMC--BM star of the field 37. In reality, Kontizas et al. and Blanco \&
McCarthy have 30 stars in common in the field 37. 

\section{Finding charts versus automatic blind match}

\begin{figure}
  \resizebox{\hsize}{!}{\includegraphics{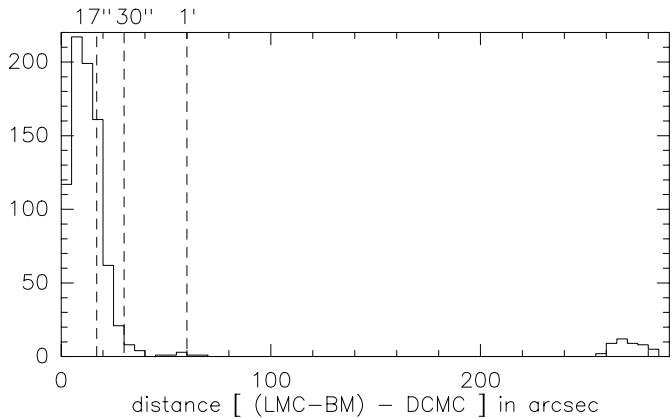}}
  \caption{Distribution of distances between LMC--BM and DCMC positions
as derived from the cross--identifications based on finding charts.}
  \label{f1}
\end{figure}

   The case of the Blanco \& McCarthy field 37 may be considered as 
an accident. However, Kontizas et al. find a surprisingly large number
of random associations in this field.
It does occur that two LMC--BM C stars are separated by less than 
$1\arcmin$, and sometimes by less than $30\arcsec$. Moreover, for various
reasons discussed in both catalogues, both surveys are incomplete, 
especially the one of Kontizas et al. in the most crowded regions (see 
their Sect.~5.3 and the field 33 in Table 1). 
It follows that to find a carbon star in both catalogues
separated by less than $1\arcmin$ does not necessarily mean that they are
the same star. From the example of the field 37 we thus expected to find
some misidentifications in the other fields.

   We have cross--identified the 849 C stars listed in the Blanco \& McCarthy
catalogue with the Kontizas et al. and 
the DCMC (Cioni et al. 2000) catalogues,
using the finding charts. We have proceeded in the following way. We use
the CDS Aladin facility. We first load a R image (ESO MAMA or AAO DSS2)
centered on the Blanco \& McCarthy coordinates. Then, comparing with their
finding chart, we mark the carbon star. Finally, using the CDS VizieR database,
we superimpose the Kontizas et al. and DCMC (or 2MASS or GSC2.2) catalogues
on the digitized image. As in general these catalogues have good
coordinates ($\sim 1\arcsec$ accurate), there is usually no doubt on
identifying the star, except in very few cases of double stars. In the
latter case, we could always identify the carbon star, by comparing
the I magnitude in Kontizas et al. and in the DCMC, or because one
star was much too blue to be a carbon star ($J-K_S < 0.5$ mag.). 

   Table 1, available electronically only at CDS, summarizes the 
cross--identifications between the Blanco \& McCarthy
and Kontizas et al. catalogues, field by field. In total, out of 849
stars, 69.0\% are re--discovered in Kontizas et al., and 99.1\%
are listed in the DCMC. The list
of the cross--identifications for individual stars is given in Table 2,
available electronically only at CDS. It gives the LMC--BM, [KDM2001],
and DCMC identifications. We found some double entries in the DCMC, which
are listed as well. Cross--identifications between the DCMC, 2MASS,
and the GSC2.2 catalogues are given in Delmotte et al. 
(2002).

   The distribution of distances between the LMC--BM 
and DCMC positions is shown in Fig.~1. According to Blanco \& McCarthy,
the accuracy of both their coordinates is expected to be 
smaller than $12\arcsec$, so
that the global error on the position is expected to be smaller
than $17\arcsec$. In fact more than a quarter of the sources have a
distance to the DCMC position larger than $17\arcsec$. The tail
of the distribution reaches $40\arcsec$. To this general distribution one
has to add 52 particular cases: (i) the 45 sources of the field 37 are
shifted by $\sim 4.5\arcmin$ as seen in Sect.~2; (ii) the 4 sources of the
field 49 are shifted by $\sim 50-55 \arcsec$; (iii) LMC--BM 6--20, 38--10,
and 42--32, are found $62$, $67$, and $48\arcsec$ away from the DCMC position,
respectively. One thus may say that the Blanco \& McCarthy positions are
erroneous for 6.1\% of their sources. Finally, we suspect that there could
be an additional error in one finding chart. 
LMC--BM 3--3, as drawn on the finding
chart, corresponds to DCMC J044614.35-675116.8 = 2MASSI J0446143-675116,
with $I-J\simeq 0.5$ and $J-K_S\simeq 0.35$ mag. This source
seems to be much too blue to be a C star. About $40\arcsec$
away, there is a C star, [KDM2001] 457. We thus suspect that [KDM2001] 457
and LMC--BM 3--3 are in fact the same star while Blanco \& McCarthy would have
indicated the wrong star on the finding chart.

   To cross--identify 849 objects by looking at the finding charts is not
the most pleasing work one could imagine, nor is it the technique one
would first think of to cross--identify catalogues. By default one would
first make a digitized version of the old Blanco \& McCarthy catalogue,
and match it automatically with that of Kontizas et al. To compare
with the results derived from the finding charts, we have also matched
blindly both catalogues. Without knowing the distribution in Fig.~1,
and being a little cautious, one would most likely use a search
radius of $30\arcsec$. We finally matched both catalogues using two
radii, $30\arcsec$ and $1\arcmin$. In case of multiple associations,
we strictly keep the closest one, checking on both entire catalogues.
The results field by field are listed in Table 1. 
Compared to the finding chart method, an automatic match leads
to 11 (1.3\%) and 20 (2.4\%) misidentifications using a radius
of $30\arcsec$ and $1\arcmin$, respectively. Among these misidentifications,
only 30\% are due to erroneous LMC--BM coordinates, all the others
are due to the presence of two close carbon stars. The number of
missed cross--identifications amounts to 5.4\% using $30\arcsec$,
and 3.3\% using $1\arcmin$. For both search radii the total number
of errors is about 6\%. This is far
from being negligible. Misidentifications are especially problematic
for individual sources. It could lead to discover strange variables,
and derive incorrect physical parameters. Missed cross--identifications
lead to a loss of information on individual sources. Both are problematic
for statistical purposes because some sources would disappear, or conversely
would be counted twice.

   In their Table 3, Kontizas et al. give cross--identifications between their
catalogue and that of Blanco \& McCarthy. According to their Sect.~5.3, they
have matched both catalogues. 
Their results are summarized in the last 3 columns
of Table 1. They find a match for 3 out of 4 LMC--BM sources of the field 49,
for LMC--BM 6--20, but not for LMC--BM 38--10 (see above),
so most likely they used an association radius of about $1\arcmin$.
However, there are some differences between their results and our match
with the same radius. 
In the field 37, we get only 7 misidentifications, while they list 13. 
The most likely reason is that, in case of multiple associations, they have 
kept the closest one field by field rather than on the entire catalogue.
Out of the 6 matches that we do not find in the field 37, one source is a 
little more than $1\arcmin$ away from 
the [KDM2001] position, and the 5 others have
a closer association in the field 42 (see also Sect.~2). The other difference
is the larger number of missed cross--identifications in the list of Kontizas
et al. than found in our match. According to Morgan (private communication)
they are due to human errors when compiling the final Table. In total,
the number of misidentifications and missed cross--identifications
amounts to 2.7 and 7.9\%, respectively, in the Table 3 of Kontizas et al.

   Finally, out of curiosity, 
we have also matched the Blanco \& McCarthy and
DCMC catalogues, using a search radius of $30\arcsec$, and taking
into account some redundant entries in the DCMC which are correctly seen as
the same source. Of course, the
density of sources in the DCMC (or in 2MASS) is such that it would not
be reasonable to just match both catalogues on the basis of the coordinates
only. We have then added some colour criteria. Almost all LMC--BM stars
have $I-J>1.2$ and $J-K_S>0.8$. The last colour criterion is confirmed by
the Fig.5 of Kontizas et al. Even applying this, a match between both 
catalogues leads to \dots a disaster, with 24.8\% of misidentifications, 
and 2.5\% of missed cross--identifications. To use a larger radius would not 
change much the result because only 23 LMC--BM sources do not have a match
within $30\arcsec$. 

\section{More errata \dots}

   It is not uncommon to find erroneous or missed cross--identifications,
or errors on the listed coordinates, in published catalogues.
During the course of more general cross--identifications in the Large
Magellanic Cloud, we found a few errors in various papers, 
sometimes real errors, sometimes probably misprints. 

   In the catalogue of M supergiants and suspected giants of Westerlund
et al. (1981), the listed coordinates are in general better than $5\arcsec$,
with a tail in the distribution up to about $20\arcsec$. We however
found 5 stars in the list of the M supergiants with erroneous coordinates
(i.e. in disagreement by more than $1\arcmin$ with the finding
chart location).
These sources are: WOH S 66, S 151, S 156, S 199, and S 212.
Like in the case of Kontizas et al. with the Blanco \& McCarthy field
37, we find that the erroneous coordinates of Westerlund et al.
have led to at least one misidentification in the
literature. We have checked on finding charts that WOH S 66, SP77 30--6
(Sanduleak \& Philip 1977), RM 1--45 (Rebeirot et al. 1983), 
and DCMC J045421.73--684524.1 are the same star. Remarkably,
the coordinates listed by Westerlund et al. for WOH S 66 are in excellent
agreement with those of Hughes (1989) for SHV 0454257--684856, alias
DCMC J045414.33--684414.2. It follows that WOH S 66 was
misidentified with SHV 0454257--684856 in Loup et
al. (1997). The IRAS source IRAS 04544--6849, LI--LMC 153
(Schwering \& Israel 1990), is more likely 
associated to the SHV star than to the M supergiant. For sure, the object 
observed with ISO in Trams et al. (1999) was the Long Period Variable
(SHV) star and not WOH S 66. 

   Old catalogues, though in principle checking carefully the 
cross--identifications on finding charts, do not all escape the problem. 
For instance Sanduleak \& Philip (1977) have erroneously identified 
SP77 46--59 with HV 2650, while the SP77 star is actually HV 996. Similarly,
SP77 51--7 is HV 5916, while Sanduleak \& Philip
give HV 591. This latter case is probably a misprint.

   The Blanco \& McCarthy catalogue has a few mistakes as well. 
In particular, it associates twice SP77 30--20, once with LMC--BM 9--13, 
and once with LMC--BM 9--23. A careful check of the finding charts shows that 
LMC--BM 9--13 is not associated to any SP77 star, that SP77 30--20
is LMC--BM 9--18, and that LMC--BM 9--23 is SP77 30--21. 
The last case is very likely a misprint. Blanco \& McCarthy also forgot
some associations. In particular, their fields 16 and 15 (``Bar West'', 
published earlier by Blanco et al. 1980, acronym BMB) overlap.
They provide 15 cross--identifications between both fields. Detailed
checks show that they missed three additional cross--identifications:
BMB--BW 37 = LMC--BM 16--16, BMB--BW 38 = LMC--BM 16--20, and 
BMB--BW 54 = LMC--BM 16--26. 

\section{Conclusions}

In practice, depending on the quality of the finding charts and on problems 
arising, and using modern facilities like Aladin at CDS, 
it is possible to check 20 to 50 finding charts per day, including
lunch and coffee breaks. To use finding charts to cross--identify the DCMC, 
2MASS, and GSC2.2 catalogues, each containing a few million stars towards 
the LMC only, one would need an army of Benedictine monks working day 
and night over 10 years. It is fortunately not required either because, 
in general, the three catalogues
provide coordinates more accurate than $1\arcsec$. 
Thus, a match based on coordinates, plus some additional validation 
criteria, allows one to reach an acceptable error 
rate (Delmotte et al. 2002).

   On the other hand, not all astronomical objects have good coordinates
yet, especially those discovered in old catalogues. Most of them have been
reobserved in modern catalogues, however the observations of these modern
catalogues do not necessarily allow to determine the nature of the object.
Thus it is important to keep our knowledge, and then to cross--identify
properly old and modern catalogues. For instance, many planetary nebulae
discovered in the Magellanic Clouds have been detected by 2MASS. But
from the 2MASS data, it is impossible to set any selection criteria to
find planetary nebulae. Most of them just look like faint blue stars,
or overlap with the large population of RGB stars, like millions of others
sources. In such a case it is obvious that to cross--identify
old catalogues of planetary nebulae with 2MASS must be done with the
finding charts. Carbon stars are a better case because
they are red objects which are much less numerous than faint blue stars.
However, as shown in this Research Note, even this is not so straightforward.
Misidentifications in the literature are also a problem for
compilation databases such as SIMBAD and NED. Matching procedures are
continuously improved, but case by case examination by an expert
often remains necessary, as demonstrated in this Note. Among others,
the contribution of B. Skiff, in this respect, is specifically 
acknowledged by the CDS.

   To conclude, if one of the catalogues has poor coordinates (accuracy worse
than $5\arcsec$), if one is not looking at an empty region of the sky, if the
objects do not have extraordinary colours or physical properties, and are
not shining like the lighthouse of Alexandria, there is no other
way to make cross--identifications than to go back to the finding charts.
Yes, it is time--consuming. 

\begin{acknowledgements}

We are very grateful to the referee Dr. H.J. Habing for his helpful
comments and suggestions. We would like to thank Dr. D.H. Morgan to have
discussed and clarified some points during the submission process.
This research has made use of the Simbad database,
operated at CDS, Strasbourg, France.
This publication makes use of data products from 
the Deep Near Infrared Survey which is the result of a joint effort involving
human and financial contributions of several institutes mostly located
in Europe, and from the Two All Micron Sky Survey,
which is a joint project of the University of Massachusetts and the Infrared
Processing and Aanalysis Center/California Institute of Technology, funded
by NASA and NSF.

\end{acknowledgements}

\end{document}